\documentclass[twoside,slac_one]{revtex4}
\usepackage{graphicx}
\usepackage{fancyhdr}
\usepackage{amsmath} 
\usepackage{bm}
\usepackage{amsxtra}
\usepackage{amssymb}
\usepackage{amsthm}
\usepackage{latexsym}
\usepackage{lscape}

\begin{document}

\title{Track Reconstruction in the NO$\mathbf{\nu}$A Experiment}

%

\author{N. Raddatz for the NO$\nu$A Collaboration}
\affiliation{School of Physics and Astronomy, University of Minnesota, Minneapolis, MN, USA}

\begin{abstract}
The NuMI Off-Axis $\nu_e$ Appearance (NO$\nu$A) experiment is a long baseline neutrino oscillation experiment using a neutrino source created from the NuMI Beamline at Fermilab. The experiment will study the oscillations of $\nu_{\mu}$ to $\nu_e$ using two functionally identical plastic, liquid scintillator filled detectors placed 14 milliradians off-axis to the NuMI beam. Charged current neutrino interactions will be used to observe the neutrino flavor from identification of the final state lepton. Reconstruction of muon charged particle tracks plays an important role in both the short term goals of detector alignment and calibration as well as longer term oscillation analyses through the identification of muon charged current events. A preliminary method of muon track recognition and track fitting based on a Kalman filter is presented.
\end{abstract}

\maketitle

\thispagestyle{fancy}
\section{The NO$\mathbf{\nu}$A Experiment}
\begin{figure*}[t]
\centering
\includegraphics[width=135mm]{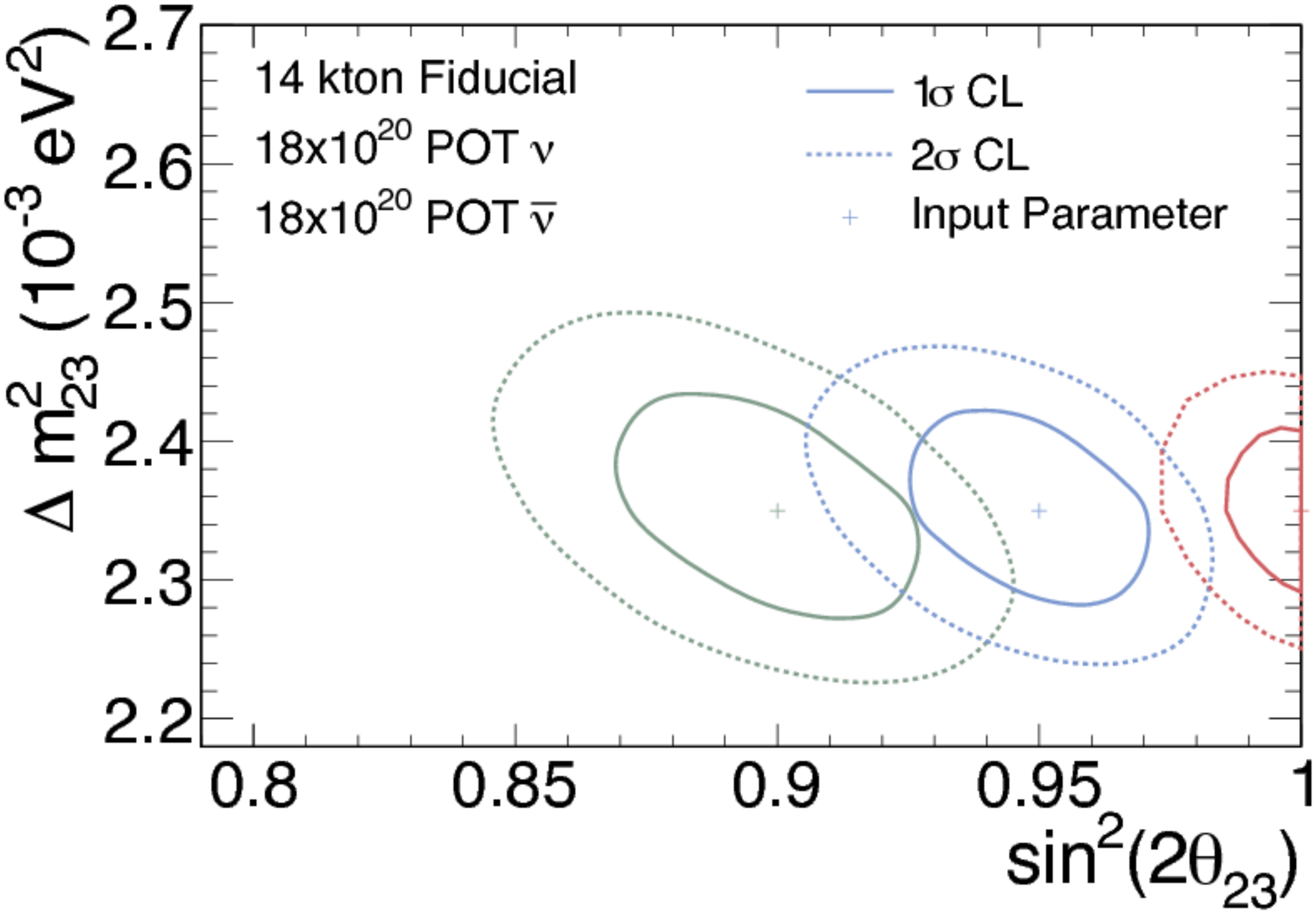}
\caption{Confidence limits on precision measurements of $\Delta m_{23}^2$ and $\sin^2(2\theta_{23})$ after 6 years of running assuming a value of $2.35\times10^{-3}$ eV$^2$ for $\Delta m_{23}^2$ and several best fit values of $\sin^2(2\theta_{23})$.} \label{sensitivity}
\end{figure*}

The NuMI\footnote{Neutrinos at the Main Injector} Off-Axis $\nu_e$ Appearence (NO$\nu$A) experiment is a long baseline neutrino oscillation experiment designed to measure the oscillation parameter $\theta_{13}$ through the observation of muon neutrinos oscillating to electron neutrinos. Depending on how large $\theta_{13}$ is, NO$\nu$A will also be able to address the neutrino mass hierarchy and charge-parity violation. In addition to these measurements, NO$\nu$A will make precision measurements of the oscillation parameters $\theta_{23}$ and $\Delta m_{23}^2$ as seen in Fig. \ref{sensitivity}. To make these measurements NO$\nu$A will use two detectors to measure the NuMI beam created at Fermilab in Batavia, IL. The NuMI beam provides a source of neutrinos by colliding 120 GeV protons with a graphite target \cite{numi}. The collisions produce primarily pions and kaons with one sign of these charged particles focused into a beam using magnetic horns. The pions and kaons decay producing muon neutrinos a majority of the time. Both NO$\nu$A detectors sit 14 milliradians off-axis to the NuMI beam and are functionally equivalent to each other with the only difference being the overall mass of each detector. The first of these detectors, the near detector, is 220 tons and is located 1 km downstream from the target, 105 m underground and measures the initial composition of the NuMI beam. The second of the the detectors, the far detector, is 14 ktons in mass, is located in Ash River, MN near the Canadian border 810 km from the near detector, and measures the oscillated composition of the NuMI beam. The off-axis location of the detectors results in a narrow energy spectrum beam of muon neutrinos around 2 GeV, which is close to the oscillation maximum for an 810 km baseline. Currently a prototype of the NO$\nu$A detectors has been constructed on the surface at Fermilab and is taking data, while the far and near detectors are scheduled to start construction this winter. A full description of the NO$\nu$A experiment is documented in the NO$\nu$A Technical Design Report \cite{TDR}.
\section{The NO$\mathbf{\nu}$A Detectors}
The NO$\nu$A detectors are designed to measure the oscillation of muon neutrinos to electron neutrinos. The primary goal of the NO$\nu$A detectors is to resolve the event topologies of electron neutrino charged current interactions, which result in an electromagnetic shower from the electron produced in this event. Additionally, the detectors must be able to reconstruct long muon tracks coming from muon neutrino charged current interactions. 

In order to be sensitive to neutrino interaction topologies the NO$\nu$A detectors are constructed in a cellular structure. Each cell is made out of reflective PVC and filled with liquid scintillator (mineral oil doped with $\sim$5\% pseudocumene), resulting in 2 GeV muons having a mean path length of 10 m. The cell dimensions for the far detector are 4 cm by 6 cm by 15 m, with each cell being 0.15 radiations lengths wide.  The cells of the near detector have the same dimensions except for the length. Each cell contains a loop of wavelength shifting fiber with both ends connected to a single pixel of an avalanche photodiode (APD). 

32 cells make up a planar detector module with each module connected to a single APD. Each APD is connected to a front end board which amplifies and shapes the APD signals creating a digitized record of measurements that is sent to the rest of the data acquisition system.

Modules are glued together to form individual planes of the detector with 12 modules per plane in the far detector and 2 or 3 modules per plane in the near detector. The detectors are constructed by gluing planes together with each detector plane rotated orthogonally to the previous plane. The detectors are placed such that the planes are oriented perpendicular to the neutrino beam. The alternating orientation of the detector planes gives two independent detector views which can be reconstructed into full three dimensional events. The PVC cellular structure forms  a ``fully active" liquid scintillator tracking calorimeter. With this design, the near detector will have a cosmic rate of 50 Hz and will see 30 neutrino events per beam spill with a 10 $\mu$sec beam spill every 1.33 s while the far detector will have a cosmic rate of 200 kHz and will see 3-4 neutrino events per day. 

In addition to the near and far detectors, a prototype detector has been constructed and is currently taking data. The prototype detector utilizes the same detector technology as the near and far detectors and is of equivalent size to the near detector. The prototype detector is located above ground approximately 1 km from the target, 110 milliradians off-axis to the beam. Currently the prototype detector is partially instrumented and taking data with a cosmic rate of 2-3 kHz and sees approximately 19 neutrino events per day if fully instrumented.
\section{Track Reconstruction and Application}
\subsection{Track Reconstruction Application}
\begin{figure*}[t]
\centering
\includegraphics[width=135mm]{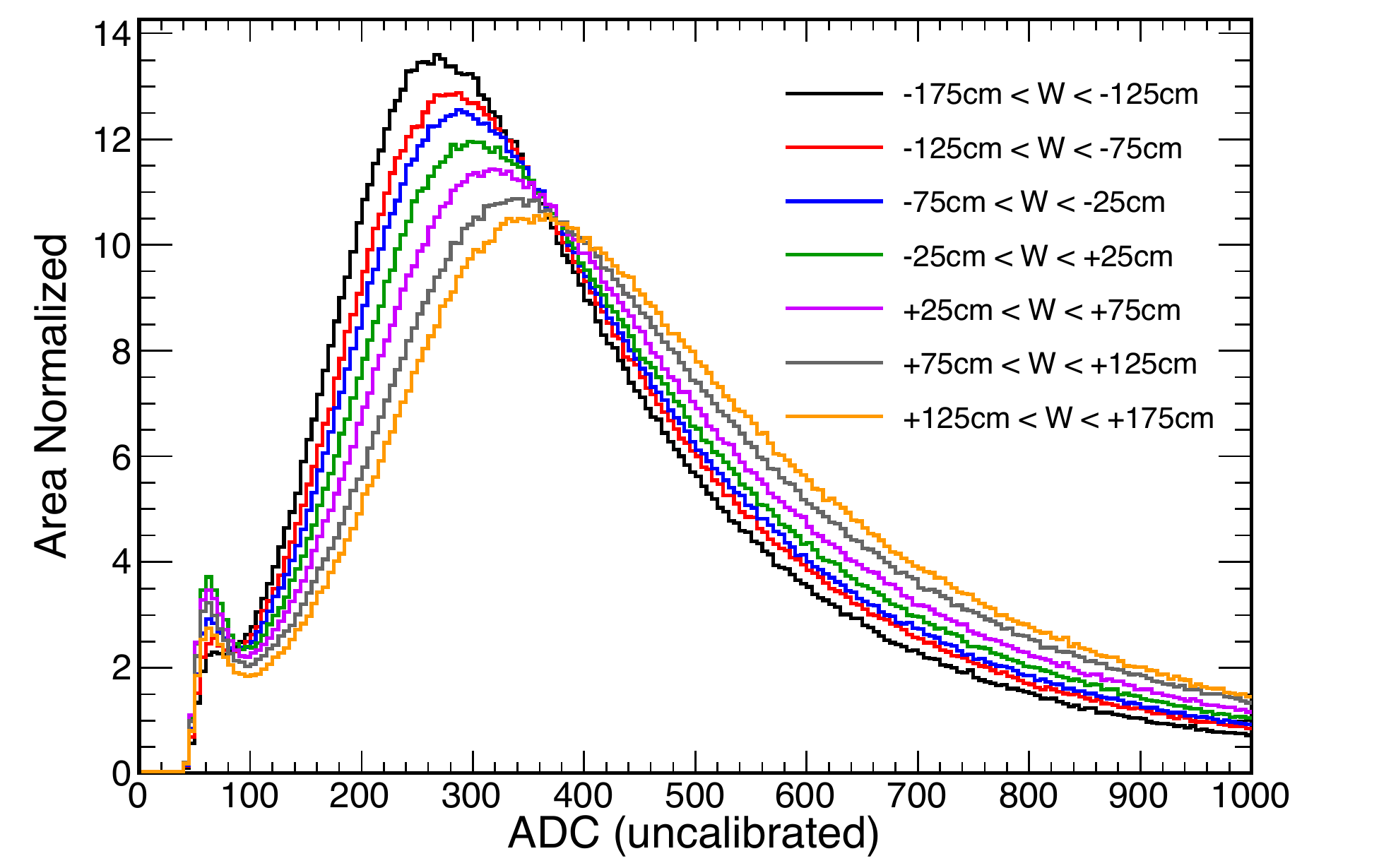}
\caption{Path length-corrected muon response for different distances from fiber end for a single example cell. W is the position of the track in the cell's long dimension with 175 cm corresponding to the end closest to the APD and -175 corresponding to the looped fiber end. The small peak close to an ADC value 75 is due to cell edge effects.} \label{adc}
\end{figure*}

\begin{figure*}[t]
\centering
\includegraphics[width=135mm]{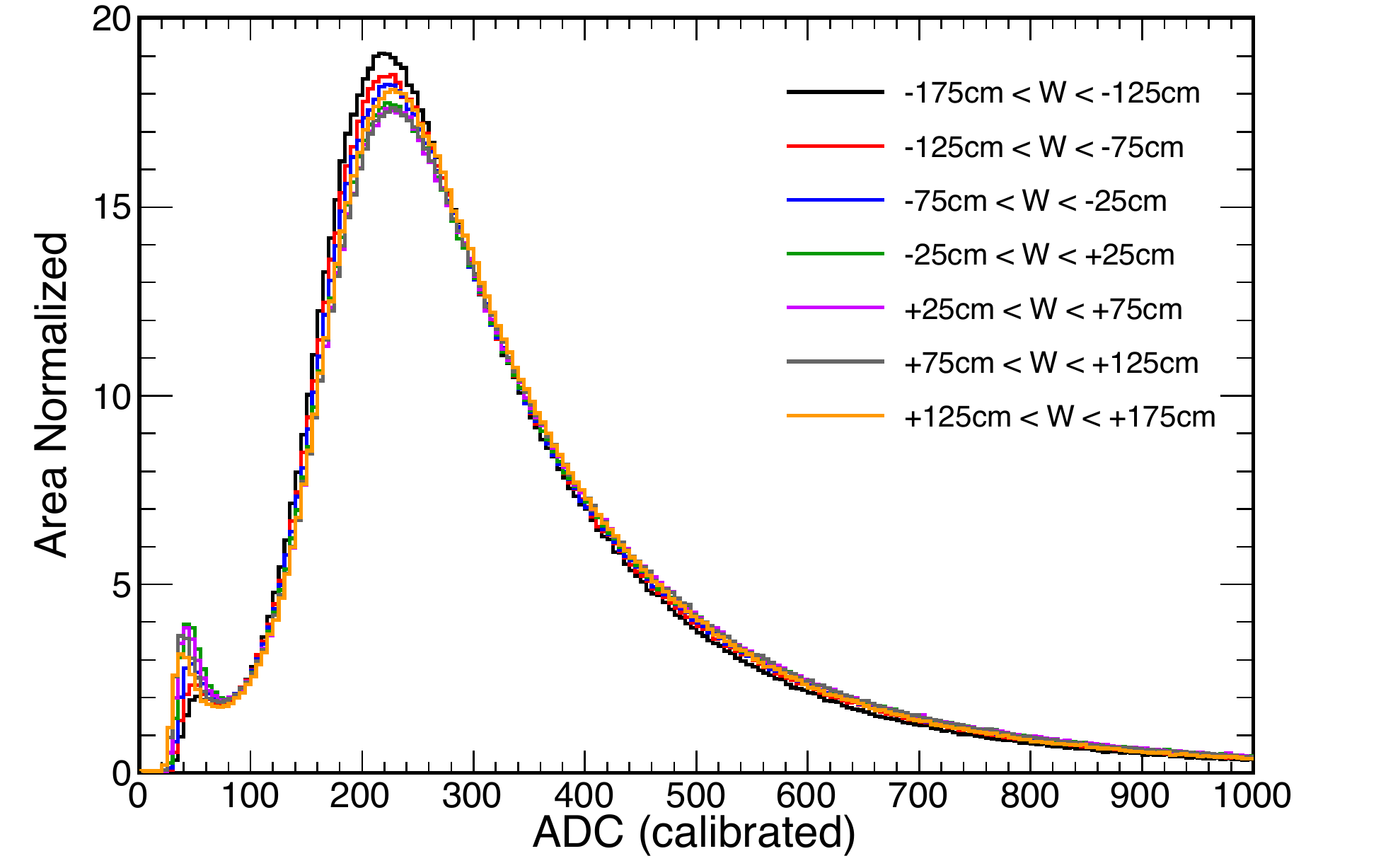}
\caption{Muon response after attenuation calibration obtained from Fig. \ref{adc}} \label{calibration}
\end{figure*}

Track reconstruction provides a general utility to help accomplish a wide range of goals in the NO$\nu$A experiment from physics analysis to monitoring detector performance. Accurate track reconstruction forms the base to determine the oscillation parameters that the NO$\nu$A experiment aims to measure. Specifically, reconstruction of muon tracks is necessary to understand the muon neutrino beam composition. In addition to its application to specific physics analyses, another example of the utility of track reconstruction is in detector calibration. Since the detectors will see a relatively large cosmic flux, reconstruction of comic muon tracks will be used to perform several levels of detector calibration. One type of calibration corrects for the differences in signal pulse heights from tracks going through cells at different distances from the APD caused by fiber attenuation. Determining this correction relies on accurate reconstruction of tracks to determine the distance from the APD readout that a track passes through the cell. This effect is shown in Fig. \ref{adc} and its correction Fig. \ref{calibration}. Another type of calibration corrects for any cell to cell differences in measured pulse heights of reconstructed cosmic ray tracks. Finally, an absolute energy calibration will be performed based on Michel electrons and the stopping power of stopped muons.
\subsection{Reconstruction Method}
Several track reconstruction methods for the NO$\nu$A detectors have been developed to address the many tracking applications, some of which were given above. One of these methods, based on Kalman filters, will be presented. The Kalman filter approach to track reconstruction was chosen because the formalism allows for both the finding and fitting of tracks in one process. Also, it can find multiple tracks within a group of time correlated hits, which is necessary to separate particles coming from the same vertex. Additionally, the Kalman filter routine can be extended to allow for the proper handling of multiple scattering \cite{billoir,fruhwirth}. Currently the reconstruction has been developed for straight tracks as an approximation to the true particle tracks which show nonlinear effects due to multiple scattering. 

\begin{figure*}[t]
\centering
\includegraphics[width=135mm]{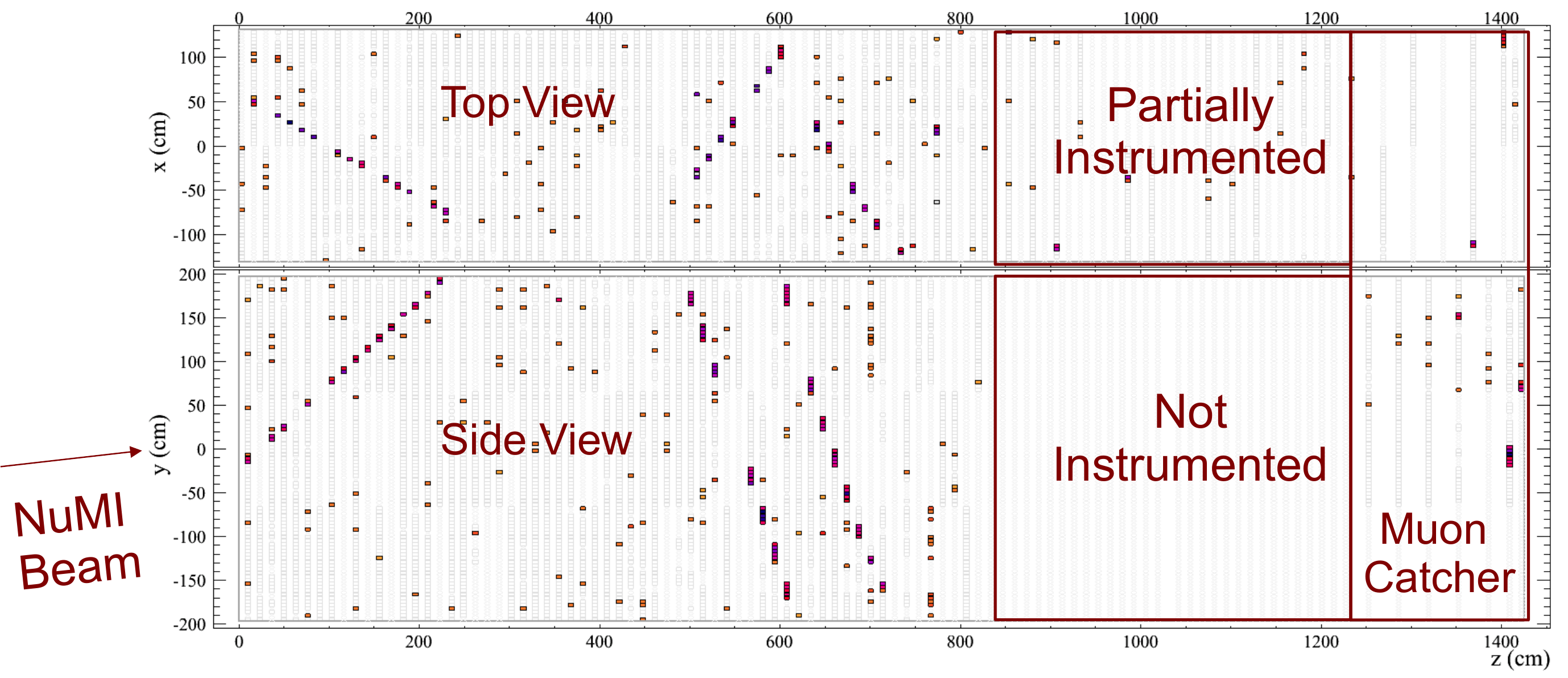}
\caption{Event Display showing cosmic data taken with the NO$\nu$A prototype detector. The Event Display shows the top and side views of the detector. The muon catcher is constructed from alternating PVC planes with planes of steel and is located on the right side of the Event Display. Each instrumented cell is shown as a light gray box.} \label{evd}
\end{figure*}

The reconstruction takes place in three steps. The first step applies a base level calibration of the hits recorded in the detector correcting for cell to cell differences in the detectors. An event display showing the calibrated hits in a full trigger window from cosmic data taken with the prototype detector is shown in Fig. \ref{evd}. 

\begin{figure*}[t]
\centering
\includegraphics[width=135mm]{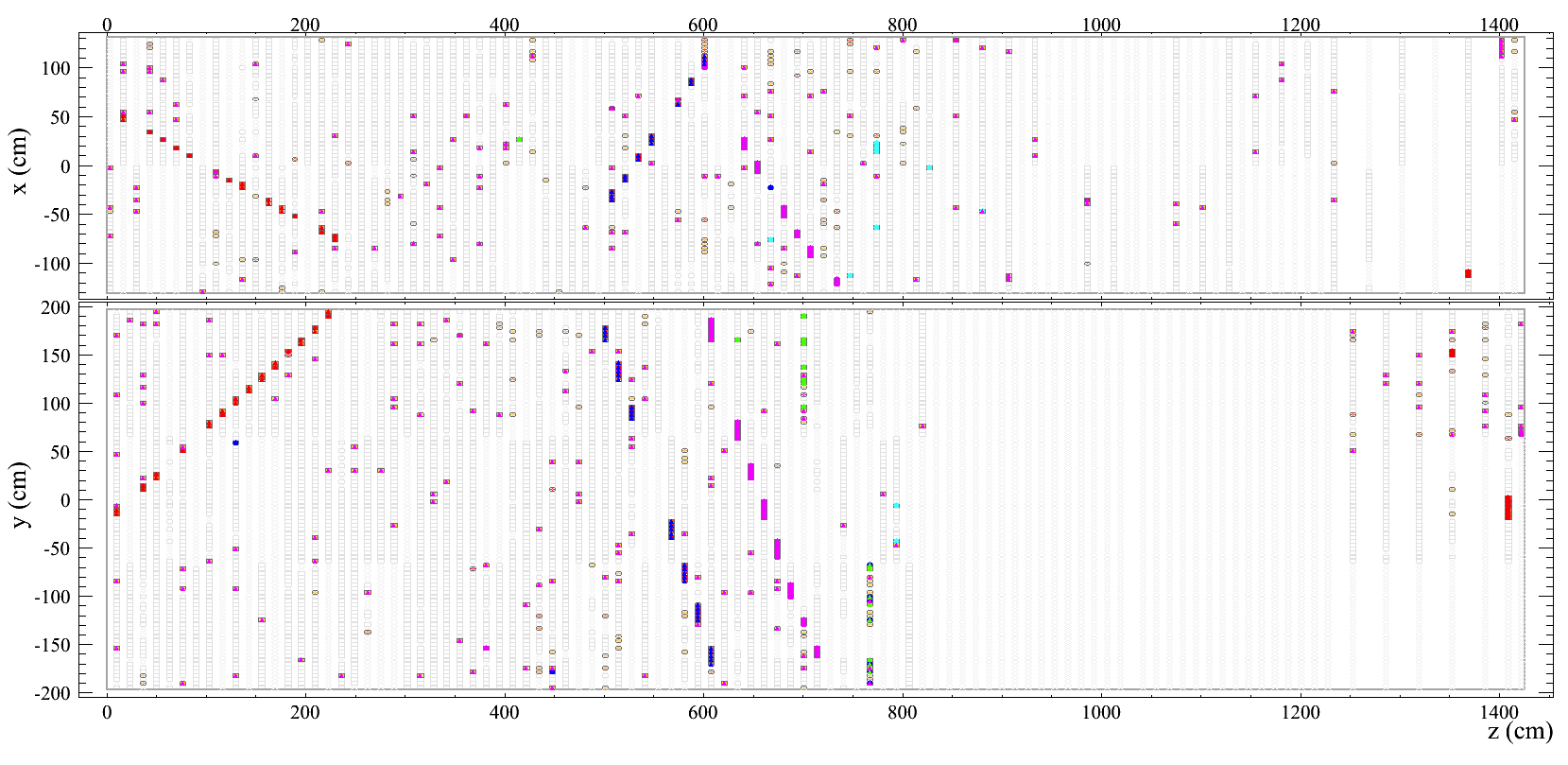}
\caption{Event Display showing time clustered hits. The color corresponds to hits belonging to the same time grouping.} \label{slicer}
\end{figure*}

The second step takes all the hits recorded in the full trigger window and clusters them into groups associated together in time by looking for a minimum level of activity in the detector without  large time gaps between hits. For reference the prototype detector's trigger window is 500 $\mu$s with groups of time clustered hits averaging to a $\sim$900 ns time duration. Figure \ref{slicer} shows the time grouping of hits from the data shown in Fig. \ref{evd}. The color indicates hits that have been grouped together. 

\begin{figure*}[t]
\centering
\includegraphics[width=135mm]{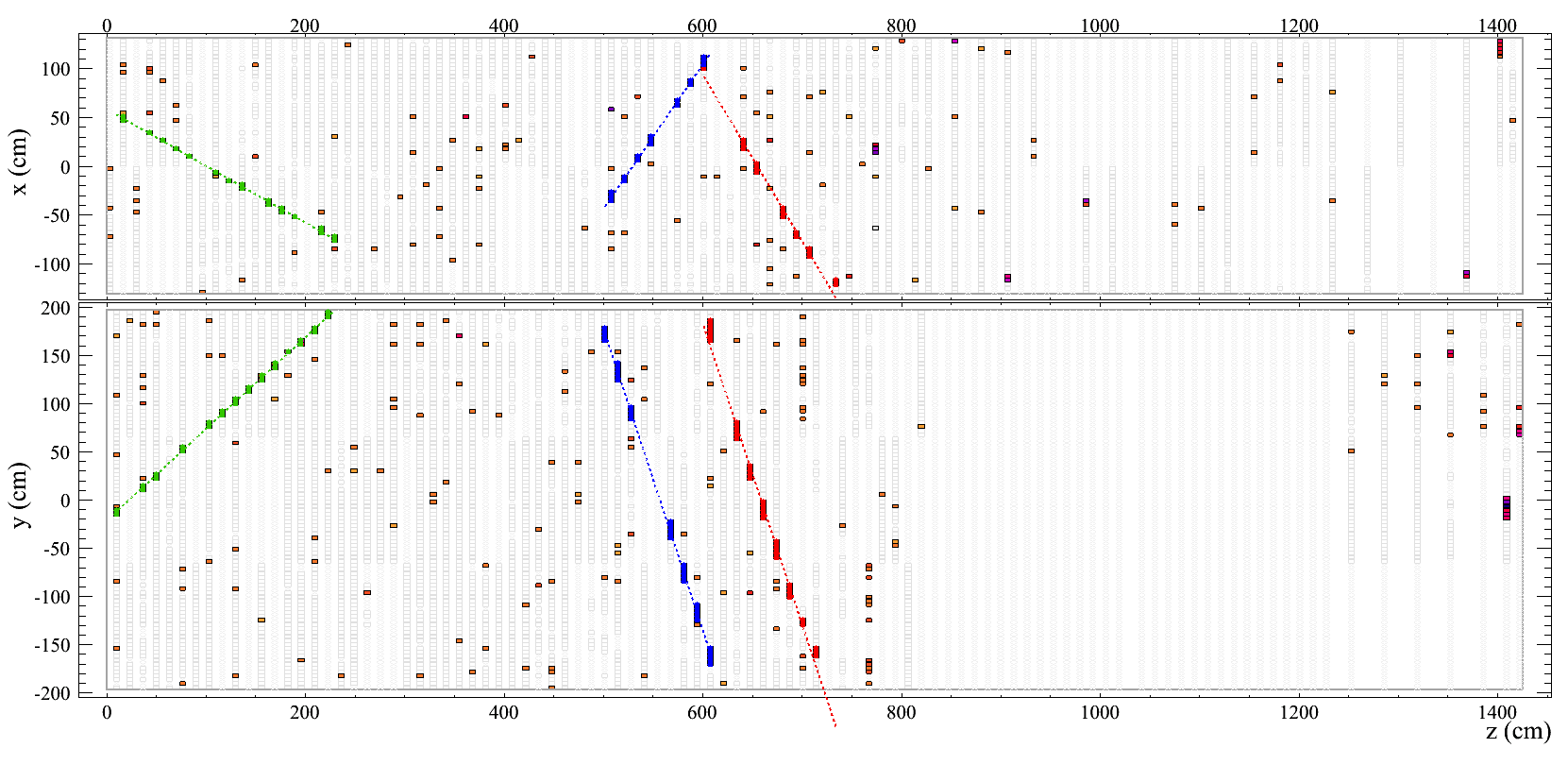}
\caption{Event Display showing fully reconstructed tracks. The color corresponds to hits belonging to the same track with the line indicating the fit of the track.} \label{tracks}
\end{figure*}

The final step of the reconstruction takes all the individual time groups of hits and applies a geometric pattern recognition routine to find tracks. The pattern recognition routine is made up of three major subroutines. The first subroutine forms track seeds by assuming that adjacent hits in each time grouped cluster belong to the same track in each independent detector view. The second subroutine then uses a Kalman filter to propagate the track seeds plane by plane through the detector adding hits to the track that are consistent with the track. The consistency of a hit is determined based on the change in $\chi^2$ of the track fit from the inclusion of the hit. The track fit is updated after the addition of any hit using the Kalman filter to perform a weighted average fit of the track to the hits. The third subroutine then takes all the tracks found in each independent view and matches them together forming a three dimensional reconstructed track. The reconstruction method requires that each track passes through 3 planes and has at least 4 hits in each view. Figure \ref{tracks} shows the fully reconstructed tracks found in the data shown in Fig. \ref{evd}. The colored hits now indicate hits that belong to the same track with the line showing the track fit.
\subsection{Preliminary Results}
\begin{figure*}[t]
\centering
\includegraphics[width=135mm]{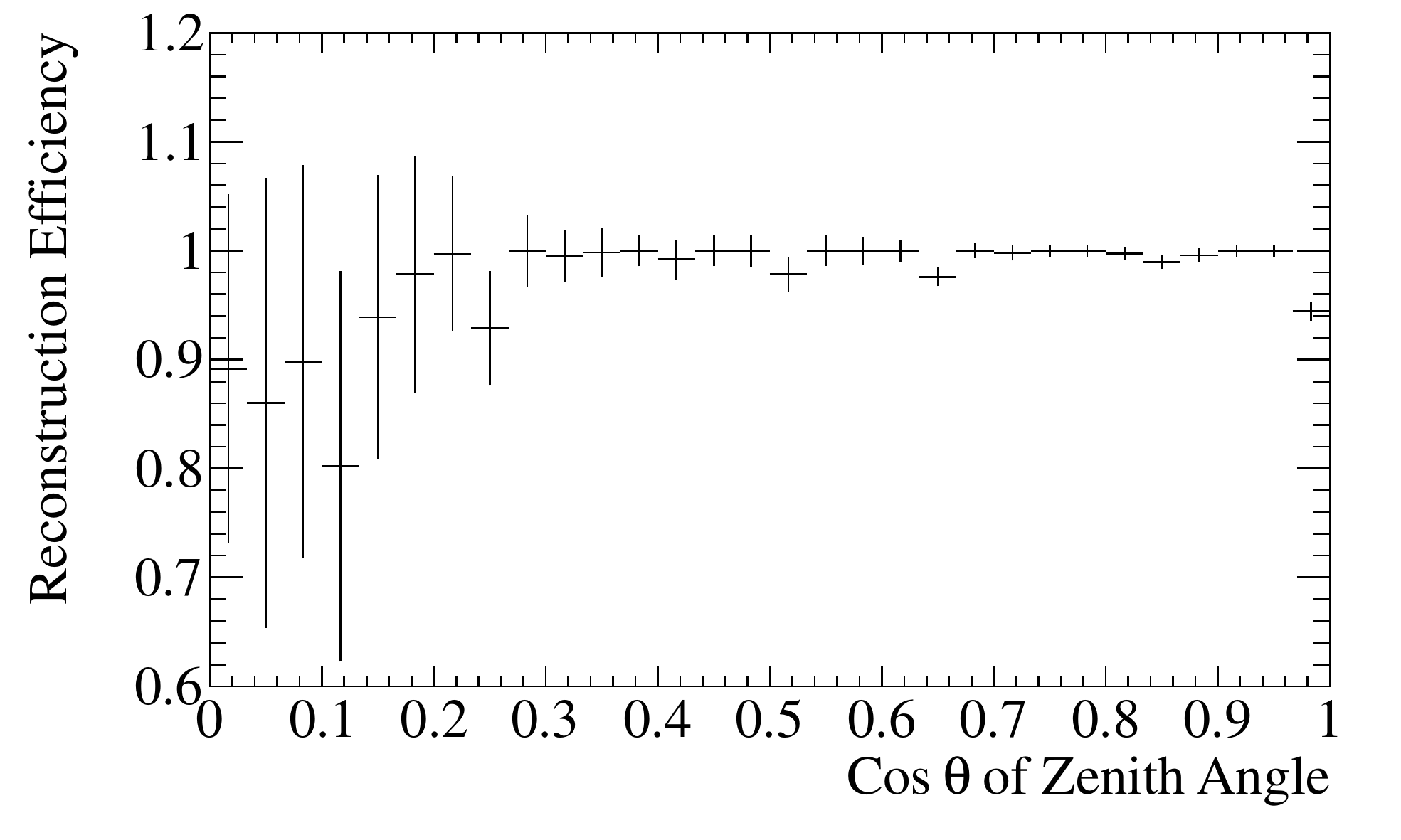}
\caption{Preliminary reconstruction efficiency of simulated cosmic ray tracks as a function of the zenith angle of the true tracks.} \label{coseff}
\end{figure*}

\begin{figure*}[t]
\centering
\includegraphics[width=135mm]{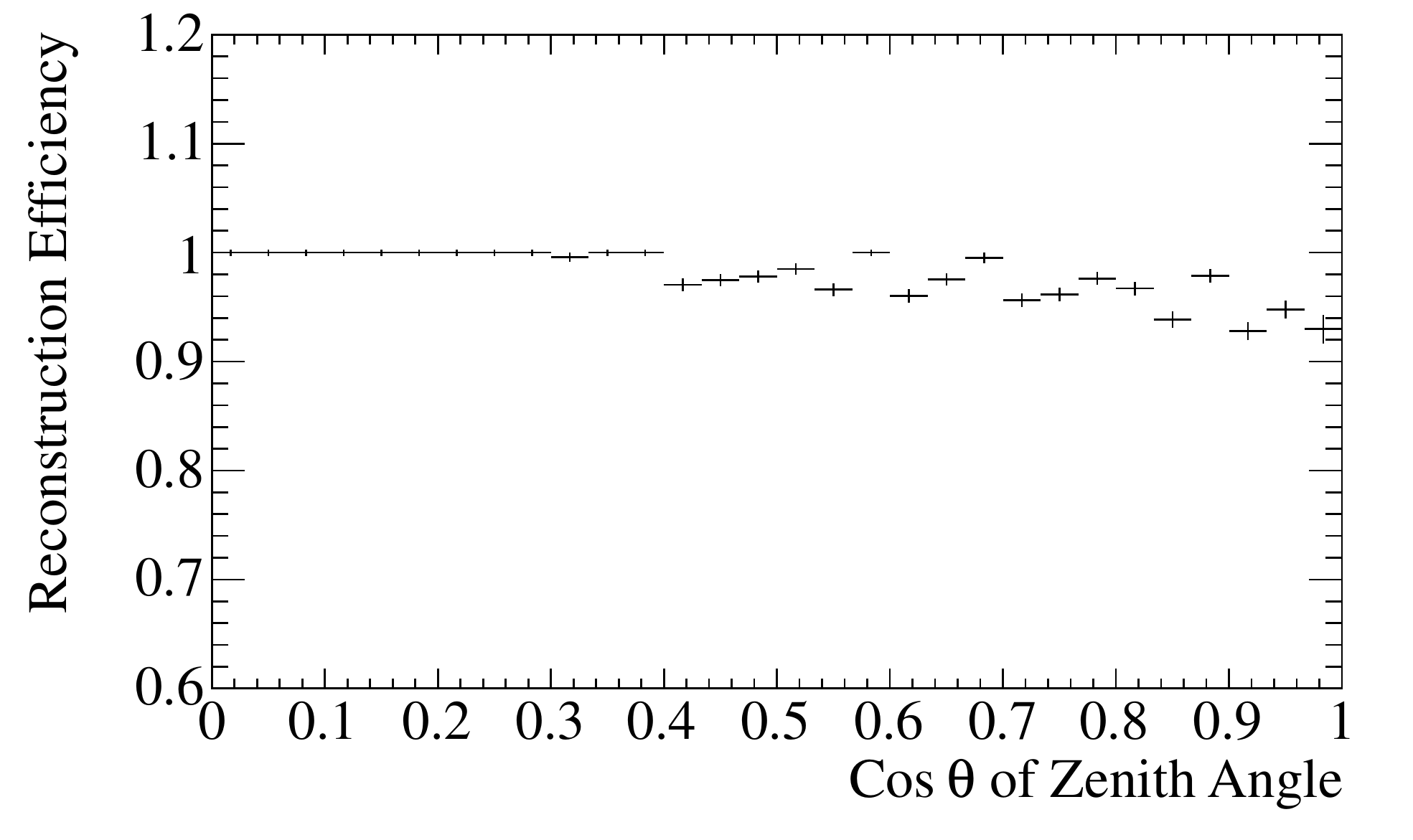}
\caption{Preliminary reconstruction efficiency of simulated uniformly distributed single particle 2 GeV muon tracks as a function of the zenith angle of the true tracks.} \label{singlepart}
\end{figure*}

The reconstruction method described above has been applied to Monte Carlo simulation and cosmic data from the prototype detector for validation. A preliminary evaluation of the reconstruction efficiency as a function of zenith angle based on Monte Carlo cosmic ray simulation has been performed and is shown in Fig. \ref{coseff}.  The efficiency is defined as the fraction of tracks that where reconstructed out of the total number of tracks that pass the reconstruction requirements. At high zenith angle the statistics dominate the efficiency calculation. To ensure that the algorithm efficiently reconstructs tracks at these angles, the reconstruction efficiency of 2 GeV uniformly distributed single particle muon Monte Carlo was calculated. The result is shown in Fig. \ref{singlepart} confirming that the technique is fully efficient for the full angular range of 2 GeV muon tracks.

\begin{figure*}[t]
\centering
\includegraphics[width=135mm]{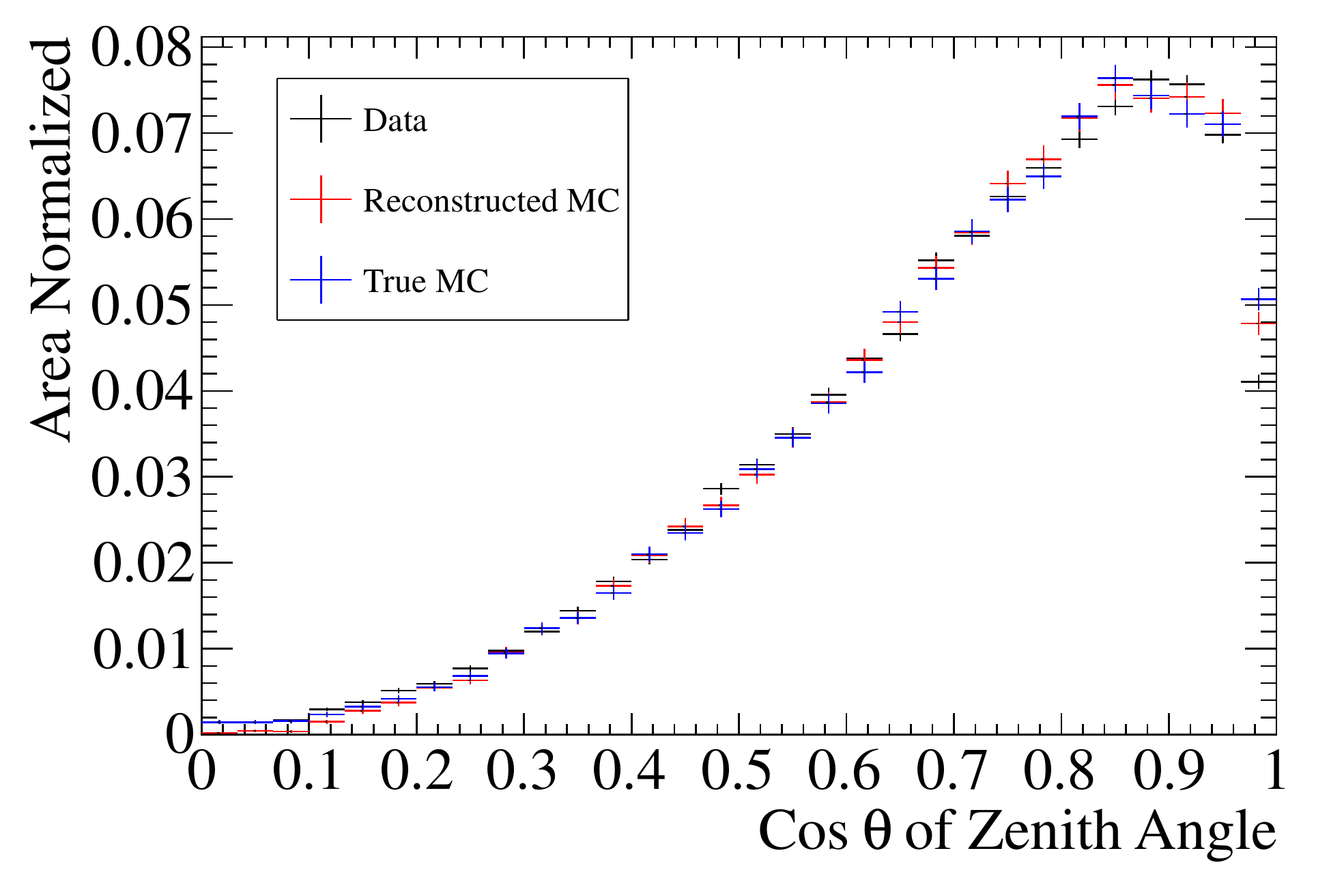}
\caption{Angular distribution of tracks from cosmic data from the NO$\nu$A prototype detector, reconstructed cosmic ray Monte Carlo tracks, and true cosmic ray Monte Carlo tracks. The fall off at low zenith angle results from the requirement that tracks pass through minimum of 3 planes.} \label{ang}
\end{figure*}

Additionally, a preliminary comparison between the cosmic ray Monte Carlo and cosmic data from the prototype detector shows the angular distribution, shown in Fig. \ref{ang}, which overall agree with each other. Some differences can be noted in comparing the prototype data to simulation as the detector is not fully instrumented or aligned where as the Monte Carlo assumes a fully instrumented, aligned detector.

\begin{figure*}[t]
\centering
\includegraphics[width=135mm]{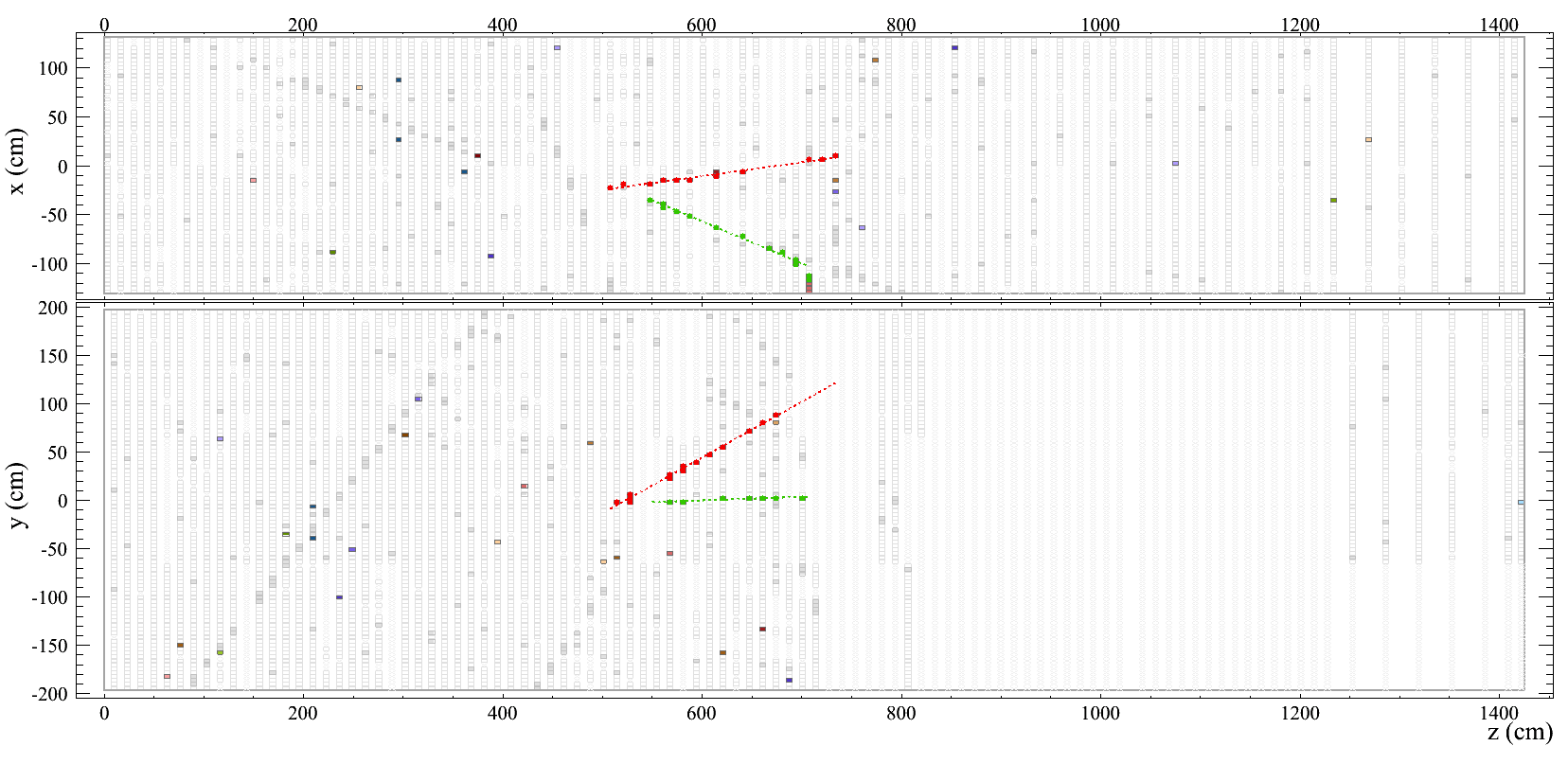}
\caption{Reconstructed candidate neutrino event from the NO$\nu$A prototype detector. Reconstruction of cosmic rays has been suppressed for clarity.} \label{candidate}
\end{figure*}

Finally, the reconstruction of tracks has been applied to candidate neutrino events that have been identified in the NO$\nu$A prototype detector data. Figure \ref{candidate} shows the reconstructed tracks from a two prong event identified as a potential neutrino event. The reconstruction has separated the hits into the two separate tracks which are identifiable by eye in the raw data. The reconstruction of the background cosmic rays in the data has been suppressed in Fig. \ref{candidate}; however, the reconstruction method does identify them allowing for separation between neutrino and background events.
\section{Summary}
The NO$\nu$A experiment requires accurate track reconstruction to accomplish both the short term goal of detector calibration as well as the long term neutrino oscillation analysis goals . Currently a track reconstruction method is in place to find and fit straight tracks in the NO$\nu$A detectors. The method is being applied to data being taken from the NO$\nu$A prototype detector and is being actively developed to improve its efficiency as well as to encompass multiple scattering effects. 


%




\bigskip 

\end{document}